\documentclass[twocolumn]{aastex631}

\newcommand{\TZO}{T\.{Z}O}
\newcommand{\TZOs}{T\.{Z}Os}

\shorttitle{Rethinking T\.{Z}O Formation in Field Binaries}
\shortauthors{Everson et al.}

\graphicspath{{./}{./}}

\begin{document}

\title{Rethinking Thorne-\.{Z}ytkow Object Formation: Assembly via Common Envelope in Field Binaries }

\correspondingauthor{R. W. Everson}
\email{rosa@ucsc.edu}

\author[0000-0001-5256-3620]{Rosa Wallace Everson}
\altaffiliation{NSF Graduate Research Fellow}
\affiliation{Department of Astronomy \& Astrophysics, University of California, Santa Cruz, CA 95064, USA}
\affiliation{Niels Bohr Institute, University of Copenhagen, Blegdamsvej 17, 2100 Copenhagen, Denmark}

\author[0000-0002-3472-2453]{Tenley Hutchinson-Smith}
\affiliation{Department of Astronomy \& Astrophysics, University of California, Santa Cruz, CA 95064, USA}
\affiliation{Niels Bohr Institute, University of Copenhagen, Blegdamsvej 17, 2100 Copenhagen, Denmark}

\author[0000-0003-1817-3586]{Alejandro Vigna-G\'{o}mez}
\affiliation{Max-Planck-Institut f\"{u}r Astrophysik, Karl-Schwarzschild-Str. 1, 85748 Garching, Germany}

\author[0000-0003-2558-3102]{Enrico Ramirez-Ruiz}
\affiliation{Department of Astronomy \& Astrophysics, University of California, Santa Cruz, CA 95064, USA}
\affiliation{Niels Bohr Institute, University of Copenhagen, Blegdamsvej 17, 2100 Copenhagen, Denmark}

\begin{abstract}
Thorne-\.{Z}ytkow objects (T\.{Z}Os), hypothetical merger products in which a neutron star is embedded in a stellar core, are traditionally considered steady-state configurations. Their assembly, especially through dynamical channels, is not well-understood. The predominant focus in the literature has been on the observational signatures related to the evolution and long-term fate of T\.{Z}Os, with their initial formation often treated as a given. However, the foundational calculations supporting the existence of T\.{Z}Os assume non-rotating spherically-symmetric initial conditions that we find to be inconsistent with a binary merger scenario. In this work, we explore the implications of post-merger dynamics in T\.{Z}O formation scenarios with field binary progenitors, specifically the role that angular momentum transport during the common envelope phase plays in constraining possible merger products, using the tools of stellar evolution and three-dimensional hydrodynamics. We also propose an alternative steady-state outcome for these mergers: the thin-envelope T\.{Z}O, an equilibrium solution consisting of a low-mass spherical envelope supported by the accretion disk luminosity of a central stellar-mass black hole. These configurations may be of interest to upcoming time-domain surveys as  potential X-ray sources that may be preceded by a series of bright transient events.
\end{abstract}


\section{Introduction} \label{sec:introduction}
The study of interacting binaries seeks, in part, to understand how the products of multiple stellar evolution differ from those expected from single stellar evolution \citep[see][for a comprehensive review]{2024PrPNP.13404083C}. To do this, it is necessary to constrain the formation channels of many types of remnant systems, including exotic or unusual merger products. Stellar mergers can result in a range of transients, such as gamma-ray bursts and luminous fast blue optical transients \citep{2023ApJ...944...74M}, as well as unusual stars, such as the too-bright B[e] supergiant of R4 in the SMC \citep{2020ApJ...901...44W} and hypothetical Thorne-\.{Z}ytkow objects \citep[T\.{Z}O;][]{1975ApJ...199L..19T,1977ApJ...212..832T}.

A \TZO\ is described as an exotic astrophysical object that may appear to be an extended post-main sequence star, but is in fact a stellar merger product with a neutron star (NS) at its core. The classical model for \TZO\ structure includes a degenerate neutron core surrounded by an inflowing nondegenerate gas ``halo'' which transitions to a convective envelope at the ``knee,'' so-called due to the discontinuity in the density-temperature relation at this location in the models of \citet{1975ApJ...199L..19T,1977ApJ...212..832T}. We hereafter refer to a classical \TZO\ as a stable post-merger configuration that reflects this structure in support of energy generation. Though both accretion and nuclear burning contribute to the luminosity of a \TZO, for those with mass below $12-16 M_\odot$, the merger product is powered predominantly by accretion onto the NS (with shell burning limited to the confines of the halo), while for higher masses the envelope may be supported predominantly by nuclear burning that extends beyond the halo into the base of the convective envelope, in an interrupted rapid proton process \citep{1989ApJ...346..277E,1991ApJ...380..167B, 1992ApJ...386..206C, 1993MNRAS.263..817C}. Though the existence of \TZOs\ is controversial, a few observations tentatively support the possibility \citep{2014MNRAS.443L..94L,2020ApJ...901..135O,2023ApJ...943...18O}. However, the preliminary conclusions of these observations have incurred numerous critiques and rebuttals \citep{2014MNRAS.445L..36T,2016MNRAS.458L...1M,2018MNRAS.479.3101B}. For a recent review, see \citet{2022ChA&A..46..164L}.

There are thought to be several potential pathways for T\.{Z}O formation: merging of a field (i.e., isolated) binary composed of a post-main sequence star and compact object (CO) through a common envelope (CE) phase \citep{1978ApJ...222..269T,1995ApJ...445..367T,1997ApJ...478..713G,2022MNRAS.513.4802A}, direct impact of a NS with its companion due to a kick \citep{1994ApJ...423L..19L,2022MNRAS.517.4544H}, or dynamical merger within a triple system or stellar cluster \citep{1987A&A...184..164R,2022MNRAS.511.4710E}. The rates of the latter two channels are estimated to be somewhat lower than the first, with recent work by \citet{2023MNRAS.523..221G} suggesting that the rate of formation via the CE channel ($\sim 10^3$ yr$^{-1}$) outpaces that of direct impact by an order of magnitude \citep{1995MNRAS.274..485P}. Therefore the CE channel in field binaries is the focus of this work. Though it has been suggested that \TZO\ formation via CE is unlikely due to the formation of jets launched by the NS as it inspirals through the stellar envelope \citep{1993ApJ...411L..33C, 2015MNRAS.449..288P}, detailed hydrodynamic simulations by \citet{2015ApJ...798L..19M} demonstrate that accretion onto NSs may become very inefficient when a density gradient is present during the CE phase, precluding jet formation during inspiral.

Historically, much theoretical effort has gone toward describing the evolution and fate of \TZOs\ while tending to avoid modeling their formation altogether \citep[e.g.,][]{1977ApJ...212..832T,1991ApJ...380..167B,1992ApJ...386..206C,2023MNRAS.524.1692F}  except, largely, through population synthesis \citep{1995MNRAS.274..485P,2018JApA...39...21H,2022MNRAS.513.4802A}. As a result, the scope of hydrodynamical simulations of \TZO\ formation is considerably limited in comparison. The foundational calculations that supported the possibility of \TZOs\ carried the assumption of a non-rotating, spherically symmetric configuration \citep[e.g.,][]{1977ApJ...212..832T, 1991ApJ...380..167B}. In the case of formation via a CE channel, rotation cannot be ignored as the angular momentum content of the material surrounding the NS has serious implications about how accretion will take place and the impact of the type of accretion on the structure of the merger product (see discussions in Sections \ref{subsec:comparison} and \ref{sec:coredisk}). It is widely accepted that accretion disk formation spells the end to a \TZO\ \citep{1995MNRAS.274..485P} due to feedback, underscoring the importance of understanding when disk formation occurs \citep{2020ApJ...901L..24M} in regards to the lifetime and overall stability of T\.{Z}Os. 

A companion paper to this work \citep{Tenley} uses 3D hydrodynamic simulations to model \TZO\ formation, providing new insights into how formation occurs as well as key factors that impact merger outcomes. \citet{Tenley} model the future merger of the X-ray binary LMC X-4 \citep{1981ApJ...246L..21L} as an ideal T\.{Z}O progenitor, exploring the effects of merger dynamics on angular momentum content, accretion rate, and energetics. The impact of angular momentum deposition via orbital decay on the stability of the core-halo-knee structure suggests further study on the constraints of T\.{Z}O formation via CE altogether, which we present here.

In this paper we aim to discover the binary conditions under which the inspiraling CO, upon merging with the core, can subsequently accrete without forming a disk and, as a result, avoid generating feedback \citep[e.g.,][]{2006ApJ...641..961L,2009MNRAS.398.2005Z,2014ApJ...781..119P,2020ApJ...901L..24M} that would prevent the formation of a classical T\.{Z}O. However, if the feedback from the accreting CO is significant, as shown by \citet{Tenley}, a classical \TZO\ is precluded, and the merger is likely to be followed by a bright transient whose properties may depend sensitively on the orientation of the observer with respect to the plane of the merging binary. In the latter case, we expect a significant fraction of the envelope to be ejected, thus challenging one of the most commonly invoked avenues for \TZO\ formation. 

The paper is structured as follows. In Section \ref{sec:formalism}, we explore the merger pathways of field binaries that are traditionally invoked for T\.{Z}O assembly and the implied outcomes of these pathways, based on angular momentum conservation. In Section \ref{sec:coredisk}, we analyze a broad parameter space of binary merger scenarios to identify the most optimistic regime for the formation of T\.{Z}Os or similar astrophysical objects. Section \ref{sec:discussion} presents an alternative to supersede T\.{Z}Os as a distinguishable transient merger product. We summarize our findings in Section \ref{sec:conclusions}.

\section{Merger Pathways} \label{sec:formalism}

We can understand the potential formation pathways of \TZOs\ and related merger products through the global properties of their respective progenitor systems, such as stellar and core mass, separation at the time of merger onset, etc. Though field binaries are not the only potential \TZO\ progenitors, the following analysis applies only to field binaries comprised of a compact object, i.e. neutron star (NS) or stellar-mass black hole (BH), and a star which are close enough to interact during the evolution of the stellar companion. 

\subsection{Common Envelope Evolution} \label{subsec:CE}
In field binaries, the onset of a common envelope (CE) interaction begins when a stellar companion, hereafter referred to as the ``primary,'' nears the end of the main sequence and begins to expand. The stellar envelope increases in radius until it has filled its Roche lobe and its companion, hereafter referred to as the ``secondary,'' begins to accrete envelope material. A CE interaction occurs when this accretion becomes dynamically unstable and the secondary is engulfed by the envelope of the primary \citep{1976IAUS...73...75P}.

A CE configuration is a frequently invoked precursor to merger scenarios, though it can also function as a mechanism for orbital tightening (i.e., hardening) in cases where the envelope is ejected and a short period binary is formed. In order to define the parameter space in which we expect \TZOs\ to form, we must exclude scenarios in which the envelope is ejected during CE. Typically, a simple energy formalism, i.e. the $\alpha$-formalism \citep{vandenHeuvel1976,Webbink1984}, is used to discern when ejection is likely.

The $\alpha$-formalism compares the orbital energy $\Delta E_\mathrm{orb}$ deposited into the envelope by the secondary to the gravitational binding energy $E_\mathrm{bind}$ of the envelope. We define $\Delta E_\mathrm{orb}$ as

\begin{equation}
\Delta E_\mathrm{orb} = \frac{G M_\ast M_\mathrm{CO}}{2 a_\mathrm{initial}} - \frac{G M_\mathrm{enc} M_\mathrm{CO}}
{2 a_\mathrm{final}},
\end{equation}
in which $G$ is the gravitational constant, $a_\mathrm{initial}$ and $a_\mathrm{final}$ are the initial and final separation, respectively, $M_\ast$ is the total mass of the extended primary star, $M_\mathrm{CO}$ is the mass of the embedded compact object, and $M_\mathrm{enc}$ is the enclosed mass of the primary at $a_\mathrm{final}$. As our interest is in merging systems, we set $a_\mathrm{final}$ to the radius of the primary's core $R_\mathrm{core}$. We then define the gravitational binding energy of the envelope as

\begin{equation}
E_\mathrm{bind} = \int_{M_\mathrm{core}}^{M_\ast} - \frac{G M(r)}{r} dm ,
\end{equation}
in which $M_\mathrm{core}$ is the mass of the core of the primary and $M(r)$ is the enclosed primary mass within the radius $r$.

Roughly, if $\alpha \Delta E_\mathrm{orb} \geq E_\mathrm{bind}$, then CE ejection is said to be successful. It is understood that various factors impact the efficiency of the orbital energy in contributing to envelope ejection and that other possible energy reservoirs may play a role; all of these considerations are combined in the efficiency term $\alpha$. Depending on the characteristics of the system, this term has been shown to have a possible range as broad as $0.01-10$ \citep{zorotovic2010}. Typical values are below unity, but to give a conservative estimate for the number of mergers we assume the transfer of orbital energy to be perfectly efficient with $\alpha=1$ \citep[for a detailed example of stellar profile analysis for CE in massive stars comparable to that applied in the following analysis, see][]{2016A&A...596A..58K}.

\begin{figure*}[tbp]
    \figurenum{1}
    \epsscale{1.1}
    \plotone{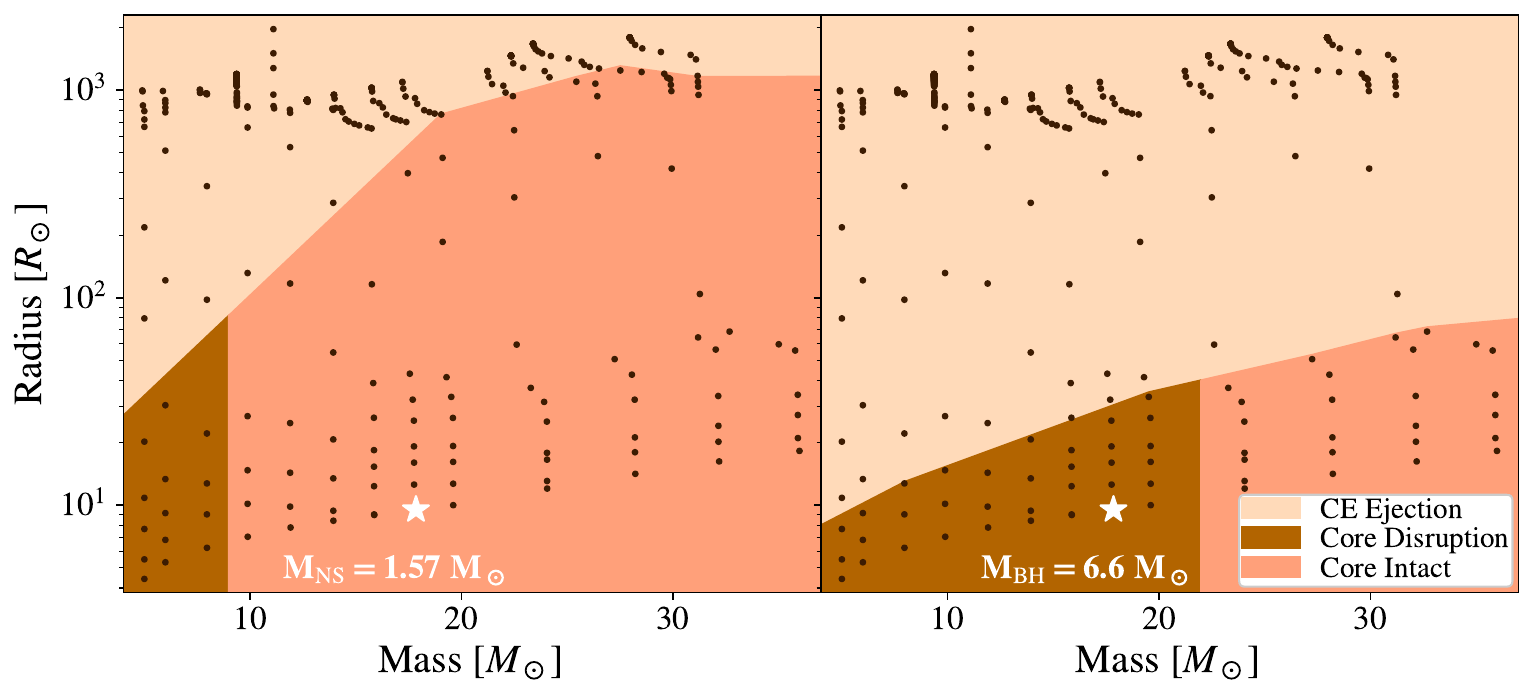}
    \caption{Outcomes of common envelope interactions between MESA models of 5-40 $M_\odot$ at solar metallicity and a typical NS (left panel) and stellar-mass BH (right panel). The MESA models (dark brown dots) evolve from the terminal-age main sequence to the tip of the giant branch, with radius as a proxy for evolutionary state. The stellar model used in the 3D merger simulations included in this work and \citet{Tenley} is indicated by the white star. Colored regions denote whether the interaction results in successful CE ejection (beige), merger with core disruption (brown), or merger with core intact (coral).}
    \label{fig:CE}
\end{figure*}

In Figure \ref{fig:CE} we show the various outcomes for a broad range of CE interactions between a typical\footnote{Typical in that their masses correspond to peaks in the COMPAS mass distribution at $1.57 M_\odot$ (NS) and $6.6 M_\odot$ (BH).} NS (left panel) or stellar-mass BH (right panel) and post-main sequence stellar companion according to this energy formalism. The compact object masses chosen correspond to the peak distribution masses of NSs and stellar-mass BHs as obtained by the COMPAS\footnote{Publicly available via the  \href{https://github.com/TeamCOMPAS/COMPAS/}{GitHub repository}.} binary population synthesis code \citep{2017NatCo...814906S,2018MNRAS.481.4009V,2022ApJS..258...34R}. We include systems with BHs here to illustrate the differences between merger products with disrupted and non-disrupted cores (Section \ref{subsec:comparison}), and to approach merger outcomes agnostically. To integrate the binding energy, we utilize a library of stellar models\footnote{The inlists utilized are available on Zenodo: \dataset[doi:10.5281/zenodo.11402979]{https://doi.org/10.5281/zenodo.11402979}.} generated with MESA v22.05.1\footnote{Results were compared to the same analysis using MESA v23.05.1 \citep{2023ApJS..265...15J} with no qualitative differences.} \citep{Paxton2011,Paxton2013,Paxton2015,Paxton2018,Paxton2019} with initial mass $5-40\, M_\odot$ at solar metallicity ($Z=0.02$) from the end of the main sequence to the maximum radius reached during the giant branch. We adopt the `Dutch' prescriptions for mass loss due to winds with scaling factor $\eta=0.8$. 

The three outcomes denoted by colored regions in Figure \ref{fig:CE} are CE ejection, merger with core disruption, and merger with core intact. The ejection scenarios must be excluded from further analysis as they do not represent \TZO\ progenitors, while the merger scenarios are explored further. We discuss the rationale for dividing the merger scenarios in the following section in accordance with Equations \ref{eqn:4} and \ref{eqn:5}. We proceed to show that the assumption that any merger between a NS and stellar companion will lead to a classical \TZO\ is unfounded, and ignores the impact of angular momentum deposition on stable \TZO\ formation.

\subsection{Comparison of Disruptive and Non-Disruptive Merger} \label{subsec:comparison}

To begin to understand the impact of angular momentum deposition on \TZO\ formation, we first focus on how the angular momentum carried by the compact object impacts the core of the primary. There are clear differences in structure between a disrupted \citep[e.g.,][]{2001ApJ...550..357Z,Jamie} and non-disrupted core \citep[e.g,][]{Tenley}, the implications of which we explore here.

Every field binary merger will spin up both the envelope material and the core of the primary \citep{2019Natur.574..211S}. Due to shocks generated from the inspiral of the compact object, the envelope will absorb most of the orbital angular momentum, but some will be deposited in the core upon merger.

\begin{figure*}[tbp]
    \figurenum{2}
    \epsscale{1.1}
    \plotone{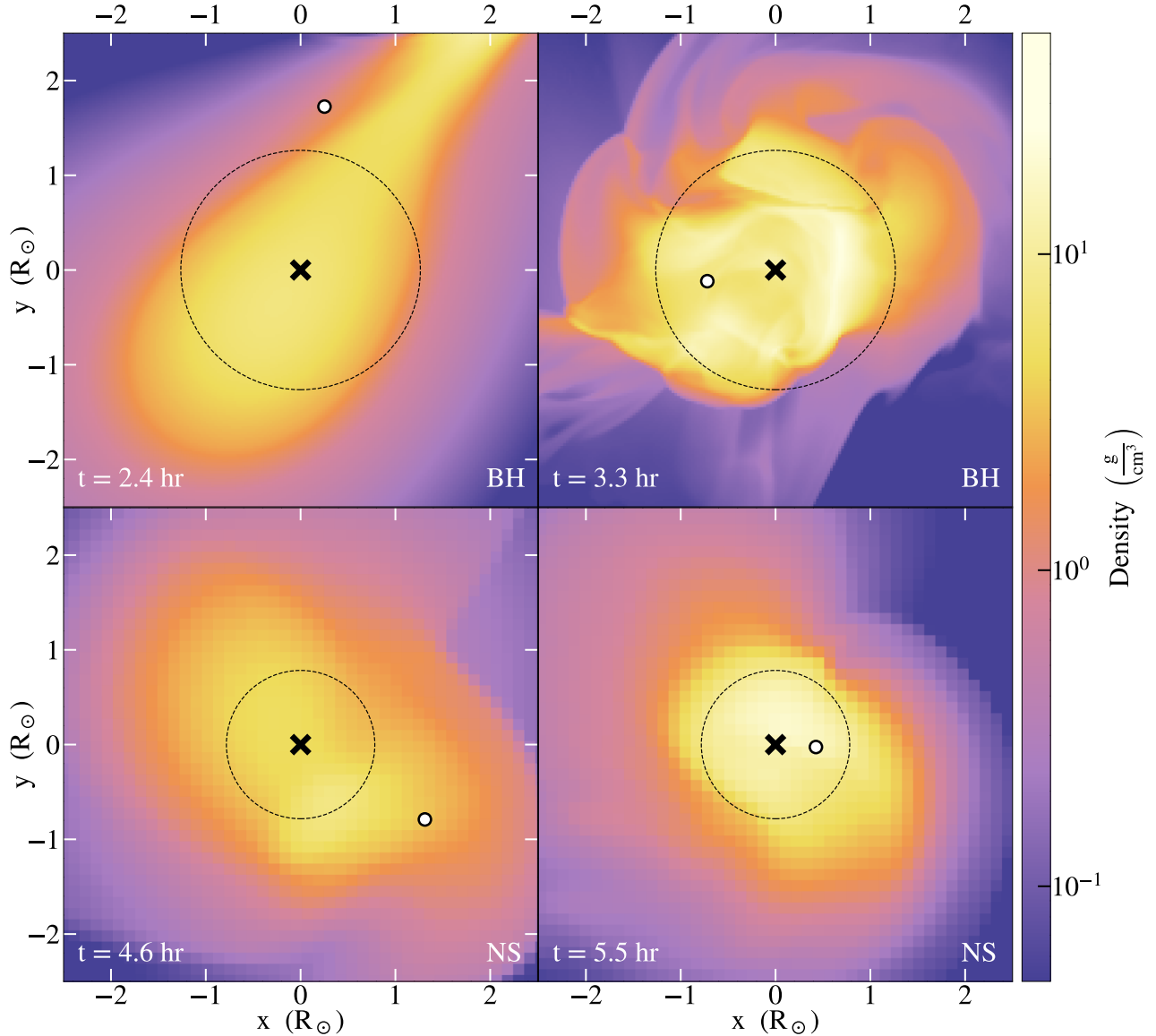}
    \caption{Comparison of simulated merger between 18 $M_\odot$ stellar model with 6.6 $M_\odot$ BH ($q_c \approx 3/2$, top panels) and 1.57 $M_\odot$ NS \citep[$q_c \approx 1/3$, bottom panels, adapted from][]{Tenley}. The black cross indicates the center of mass of the primary, the white circle with black outline indicates the position of the inspiraling compact object, and the black dashed circle indicates the tidal radius. In the top panels, note that the core (yellow to orange) is deformed as the BH approaches the tidal radius, and is then fully disrupted, indicated by the distribution of shock-heated material surrounding the center of mass. In the bottom panels, we still see deformation to a lesser degree, with the highest density material remaining centrally concentrated.}
    \label{fig:3D}
\end{figure*}

In Figure \ref{fig:3D}, we demonstrate the impact on core structure of a disruptive and non-disruptive merger by comparing the results of two representative 3D hydrodynamical simulations using the setup described in \citet{Tenley}, adapted from \citet{2020ApJ...901...44W} and \citet{Jamie}. Using the FLASH adaptive mesh refinement hydrodynamics code \citep{2000ApJS..131..273F}, we map the non-rotating MESA profile of a star of initial mass $18\ M_\odot$ at $9.5\ R_\odot$ as it leaves the main sequence, resolving the core. We introduce a point-mass compact object moving at Keplerian velocity at the limb which proceeds to plunge inward due to drag, shock-heating the envelope. Figure \ref{fig:3D} provides a zoomed-in view of the stellar core near the end of the compact object inspiral. Top panels show the case of a typical BH of mass $6.6\ M_\odot$ and bottom panels show the case of a typical NS of mass $1.57\ M_\odot$ when the compact object is approximately $0.5\ R_\odot$ from the tidal radius $R_\mathrm{tidal}$ (left) and when the compact object reaches $0.5R_\mathrm{tidal}$ (right), defined as follows:
\begin{equation}
    R_\mathrm{tidal} \approx \left( \frac{M_\mathrm{CO}}{M_\mathrm{core}} \right)^{1/3} R_\mathrm{core} , 
\end{equation}
in which $M_\mathrm{CO}$ is the mass of the compact object and $M_\mathrm{core}$ and $R_\mathrm{core}$ are the mass and radius of the core, respectively. The global parameters of these simulations are indicated by white stars in both panels of Figure \ref{fig:CE}.

Though classical \TZOs\ are formed only from NS CE events, we introduce a BH here in order to give a one-to-one comparison of the core structure of the same primary star at the same stage of evolution in both the disrupted and non-disrupted case. Utilizing a secondary with a larger mass allows us to test both cases controlling for all other factors.

Though both cores are initially deformed by the incoming compact object, the more intense shocks created by the BH completely disrupt the core while the less intense shocks of the NS allow the highest density material to remain centrally concentrated. Both cores are spun up through this process, but as long as the spin is lower than that required to form a disk, we can assume quasi-spherical accretion that is requisite to power a classical T\.{Z}O.

An accounting of the angular momentum budget as the compact object merges with the core may provide us with a parameter space of progenitor systems in which quasi-spherical accretion is possible.

\subsubsection{Disruptive Merger}
In cases in which an inspiraling compact object will disrupt the core of its stellar companion prior to merging with it, an accretion disk will be formed. 
This occurs when the tidal radius is greater than the size of the stellar core, therefore
\begin{equation} \label{eqn:4}
    \frac{R_\mathrm{tidal}}{R_\mathrm{core}}
    \approx \left( \frac{M_\mathrm{CO}}{M_\mathrm{core}} \right)^{1/3} > 1,
\end{equation}
causing material to be dynamically stripped from the core and ejected by tidal torques through the outer Lagrange points, transporting angular momentum and forming an extended centrifugally supported structure \citep{2016ApJ...823..113P}. 

Here, we define the mass ratio between the compact object and the core of the companion as
\begin{equation} \label{eqn:5}
    q_c := \frac{M_\mathrm{CO}}{M_\mathrm{core}}, 
\end{equation}
giving the criterion that for a disruptive merger, $q_c > 1$, which defines the vertical boundary between the ``Core Disruption'' and ``Core Intact'' regions of Figure \ref{fig:CE}. An accretion disk is guaranteed in this case, precluding classical T\.{Z}O formation.

Therefore we limit our remaining analysis to systems in which $q_c < 1$, as shown in the ``Core Intact'' regions of Figure \ref{fig:CE}. 

\subsubsection{Non-disruptive Merger}
\begin{figure}[tbp]
    \figurenum{3}
    \epsscale{1.15}
    \plotone{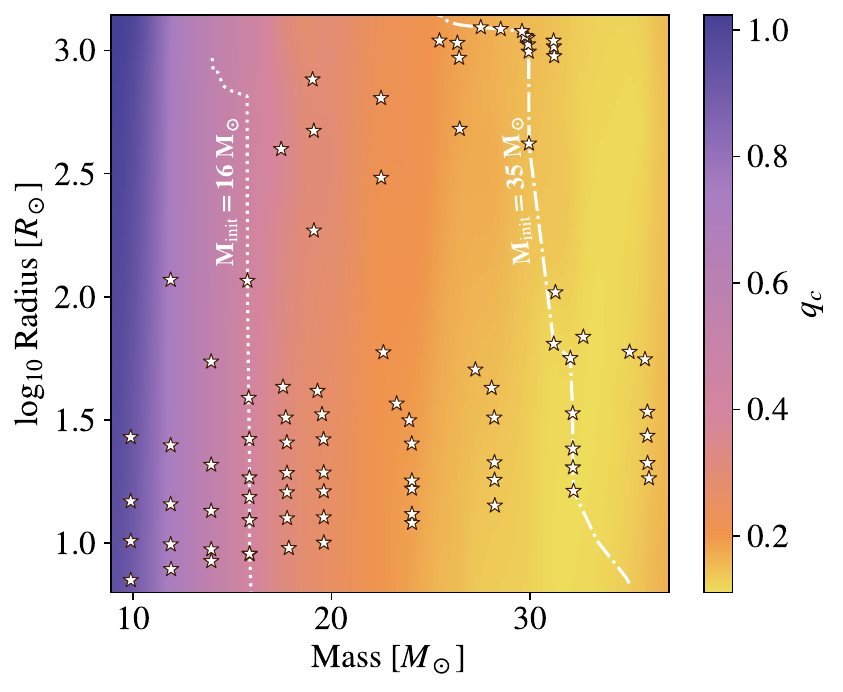}
    \caption{NS merger models in which the core remains intact (white stars), overplotted on mapped values of the mass ratio of the compact object and core: $q_c:=M_\mathrm{CO}/M_\mathrm{core}$. Stellar tracks evolve bottom to top, and slightly leftward due to wind-driven mass loss, with representative tracks shown for initial masses of $16 M_\odot$ (dotted line) and $35 M_\odot$ (dot-dashed line). Note that, although total stellar mass decreases during post-main sequence evolution, the $q_c$ values shown here are more or less constant with radius because core masses change very little within this parameter space. Complete core disruption occurs for $q_c>1$, shown on the left edge in dark purple.}
    \label{fig:qc}
\end{figure}

For typical \TZO\ progenitor systems, which is to say primaries partnered with NSs, excluding those which are likely to lead to envelope ejection or core disruption, $q_c$ values remain fairly constant. In Figure \ref{fig:qc}, we map the $q_c$ values over the relevant parameter space interpolated from the MESA library. Evolutionary tracks proceed from bottom to top as radius increases, and slightly from right to left due to wind-driven mass loss. Recalling that NS accretion during inspiral is very minimal \citep{2015ApJ...803...41M,2015ApJ...798L..19M,Tenley}, we assume a $1.57 M_\odot$ NS with constant mass, and note that although cores become more compact during evolution on the giant branch, their mass increases only slightly, if at all, within this parameter space. This allows us to simplify our approach with the assumption that for any model of a given initial mass that satisfies $q_c<1$, any profile from its evolution still satisfies this criterion in the merger regime.

Knowing the value of $q_c$, though sufficient as a disruption criterion, does not directly inform whether an accretion disk is formed. In fact, disruption of the core is not required for accretion disk formation, suggesting there may be a critical rotation rate $\Omega_\mathrm{crit}$ below which T\.{Z}O formation is possible. To derive this value, we first define the minimum specific angular momentum required to maintain the innermost stable circular orbit (isco) about the compact object as
\begin{equation} \label{eqn:6}
    j_\mathrm{isco} \approx \frac{2 G M_\mathrm{CO}}{v_\mathrm{esc}} ,
\end{equation}
in which $v_\mathrm{esc}$ is the escape velocity of the compact object. Post-merger, the core material surrounds the central compact object, and has been spun up to some degree by the end of the inspiral process. We can approximate the total angular momentum of the core as 
\begin{equation} \label{eqn:7}
    J \approx M_\mathrm{CO} \sqrt{G (M_\mathrm{CO} + M_\mathrm{core}) R_\mathrm{core}} ,
\end{equation}
assuming the limiting case in which the inspiraling compact object transfers all of its angular momentum to an initially non-rotating core. This gives a rotation rate of $\Omega = J/I_\mathrm{core}$, in which $I_\mathrm{core}$ is the moment of inertia of the core given solid-body rotation. 

\begin{figure}[tbp]
    \figurenum{4}
    \epsscale{1.15}
    \plotone{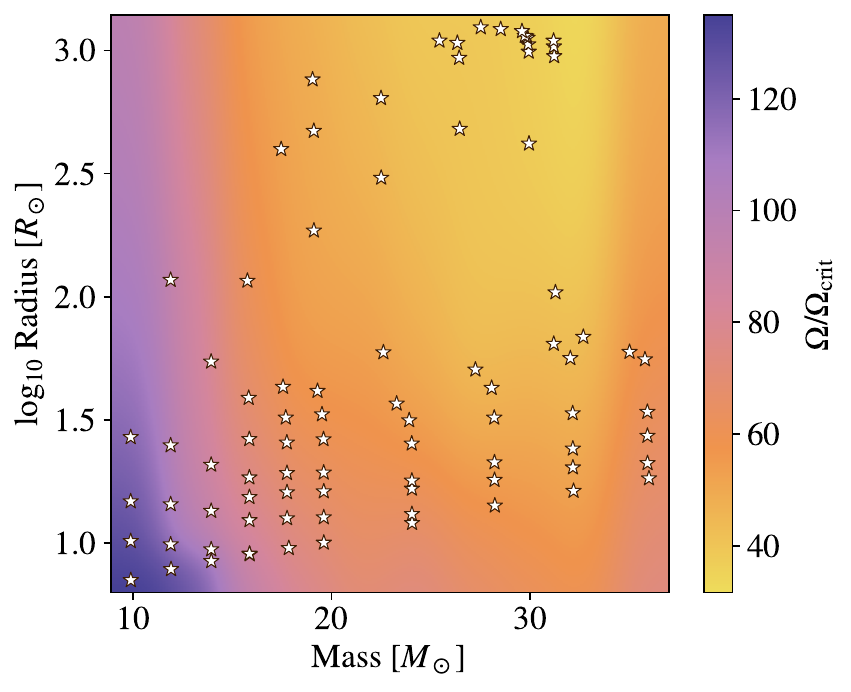}
    \caption{Minimum rotation rates for mergers with NS in which the core remains intact, normalized by the critical rotation rate for disk formation. Modeled systems are overplotted with white stars. Typical rotation rates in this parameter space are 1-2 orders of magnitude higher than $\Omega_\mathrm{crit}$, showing disk formation in the core to be inevitable upon merger.}
    \label{fig:omegacrit}
\end{figure}

In order to avoid accretion disk formation, the specific angular momentum of the core material must be less than that required to main orbit, $j_\mathrm{isco}$. Therefore, the rotation rate of the core $\Omega$ must be less than the critical rate of
\begin{equation}
    \Omega_\mathrm{crit} = \frac{j_\mathrm{isco} M_\mathrm{core}}{I_\mathrm{core}} .
\end{equation}
This provides an expression for the critical rotation rate that is dependent on the properties of the compact object and the core:
\begin{equation}
    \Omega_\mathrm{crit} = \frac{2 G M_\mathrm{CO}}{\xi v_\mathrm{esc} R_\mathrm{core}^2} ,
\end{equation}
in which $\xi = I_\mathrm{core}/M_\mathrm{core} R_\mathrm{core}^2$ parameterizes the internal structure of the core after the merger. We then have the condition for T\.{Z}O formation that
\begin{equation}
    \frac{\Omega}{\Omega_\mathrm{crit}} = \frac{v_\mathrm{esc}}{2 M_\mathrm{core}} 
    \sqrt{\frac{(M_\mathrm{CO} + M_\mathrm{core}) R_\mathrm{core}}{G}} < 1 .
\end{equation}
The values of this expression mapped over the \TZO\ progenitor systems are shown in Figure \ref{fig:omegacrit}. Throughout the relevant parameter space, the minimum of this value is $\Omega/\Omega_\mathrm{crit} \approx 36$, found only in the most extended and most massive merger models, which implies that disk formation is inevitable. Yet, perhaps disk formation is not instantaneous throughout the core: the specific angular momentum $j$ of the core material is a function of density in the core and cannot be treated as constant. We proceed to look more deeply into core structure to investigate how and where disk formation may occur.

\section{Implications of Core Structure on Disk Formation} \label{sec:coredisk}

In order to perform a broad analysis of angular momentum content in a range of progenitor cores upon merger, we use the MESA model library described in Section \ref{subsec:CE} to cover the relevant parameter space. In all cases, we use the conservative estimate that the post-merger core undergoes solid-body rotation at the rate defined by the total angular momentum content of the secondary at $R_\mathrm{core}$, neglecting any prior spin up due to shocks. 

\begin{figure}[tbp]
    \figurenum{5}
    \epsscale{1.1}
    \plotone{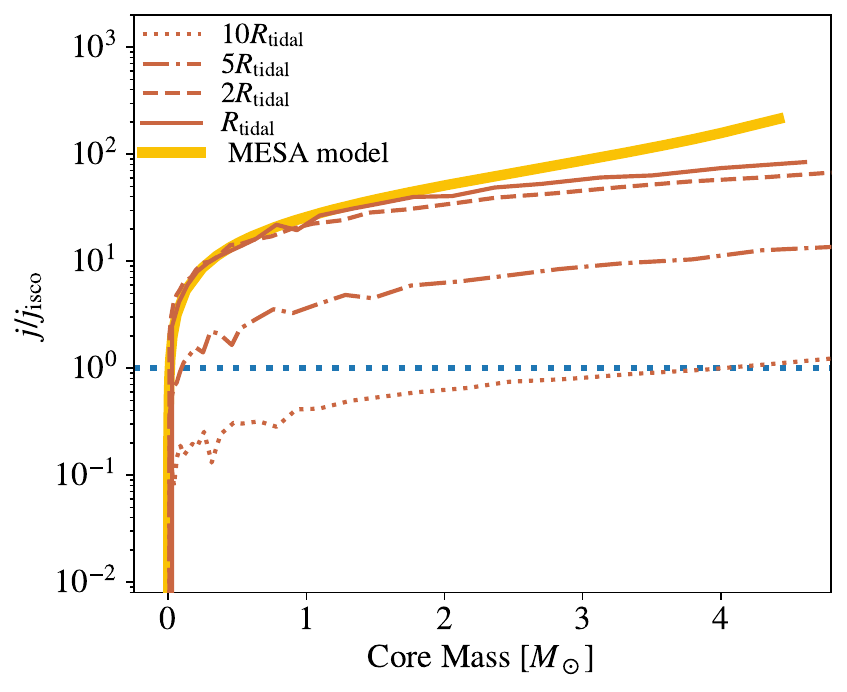}
    \caption{Comparison of normalized specific angular momentum content $j/j_\mathrm{isco}$ of the core of 18 $M_\odot$ model in FLASH (3D) and MESA (1D). The specific angular momentum ratio required for disk formation ($j/j_\mathrm{isco}>1$) is delimited by the blue dotted line. The MESA model corresponding to the FLASH simulation is shown in yellow, spun up in solid-body rotation to the total orbital angular momentum of a NS secondary at $R_\mathrm{core}$. The angular momentum from the FLASH simulation is shown in copper at different depths during inspiral, from 10 $R_\mathrm{tidal}$ to $R_\mathrm{tidal}$ (within the core, in this case). The 1D model in solid-body rotation gives a close approximation to the initial state of the core during merger, therefore we use this approach to investigate the prevalence of disk formation in cores across the parameter space during merger.
    }
    \label{fig:jcomp}
\end{figure}

In Figure \ref{fig:jcomp}, we show that this approximation is consistent with the core's angular momentum content as measured in our 3D hydrodynamic simulations at the end of inspiral. The NS case is shown for a secondary of $1.57 M_\odot$, with the specific angular momentum profiles normalized by $j_\mathrm{isco}$ as in Equation \ref{eqn:6}. Roughly, $j/j_\mathrm{isco}$ is a more detailed, structure-dependent proxy of $\Omega/\Omega_\mathrm{crit}$, in that these ratios would be equivalent if $j$ is a constant for the core. The profiles in copper are calculated from a 1D spherical average about the center of mass in the 3D FLASH simulation described in Section \ref{subsec:comparison}, shown at various radii during inspiral. Depicted in yellow is the same quantity derived from the initial mass $18\ M_\mathrm{\odot}$ MESA model that provided the stellar structure for the FLASH simulation.

We define the mass and radius of the MESA profile's stellar core using the traditional core boundary criterion of $X_\mathrm{H} = 0.1$; the appropriate definition of the core boundary in CE and merger calculations is still an area of active discussion, and we defer to that used most often in the literature as it provides a lower limit for available angular momentum while recognizing that more nuanced definitions may ultimately be more physically relevant \citep[e.g.,][]{2001A&A...369..170T,2013A&ARv..21...59I, 2020ApJ...899...77E,2022MNRAS.511.2326V}. 

We then calculate the total angular momentum (Equation \ref{eqn:7}) as the orbital angular momentum of the NS at $R_\mathrm{core}$ and derive the rate of solid-body rotation $\Omega$ with the integrated moment of inertia $I_\mathrm{core}$ from the MESA profile. The radial profile of the core combined with the rate of rotation gives the specific angular momentum profile, which we then compare to our simulation results. 

It is clear that in the simulated merger, shocks during inspiral begin to spin up the core long before the NS comes into contact with it (Figure \ref{fig:jcomp}, copper profiles from bottom to top), however this doesn't create a large departure from the 1D MESA profile (yellow). For non-disrupted cores such as this one, $R_\mathrm{tidal} < R_\mathrm{core}$, so the specific angular momentum profile at $R_\mathrm{tidal}$ gives a snapshot of the rotation as merger is occurring. The assumption of solid-body rotation in the core upon merger gives a close approximation to the rotation of the simulated core at that moment.

Most notable is how much greater than unity the profiles shown in Figure \ref{fig:jcomp} are: the blue dotted line represents the minimum value of $j$ required for disk formation about the NS, and this specific stellar model achieves this while the NS is still plunging through the envelope. Nonetheless, differences in the internal structure of the core during post-main sequence evolution will impact the shape of these profiles, further motivating the analysis of the full set of MESA models in search of a case that does not meet this minimum.

\begin{figure*}[tbp]
    \figurenum{6}
    \epsscale{1.15}
    \plotone{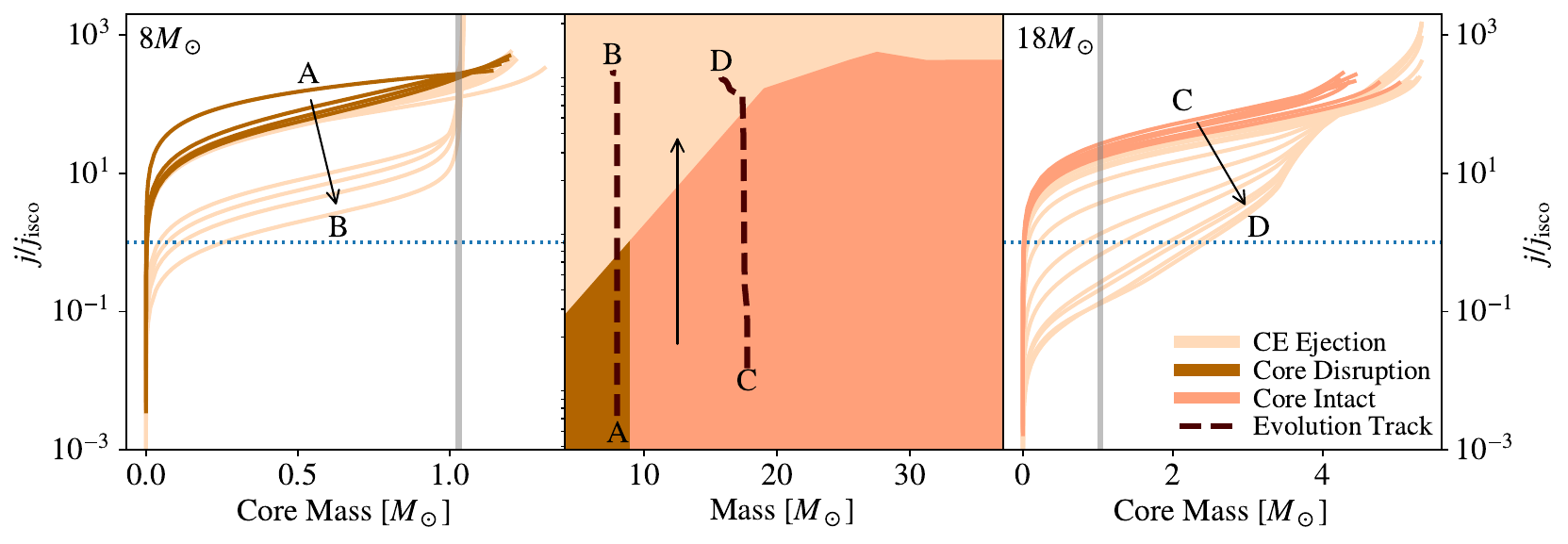}
    \caption{Normalized specific angular momentum of cores of 8 $M_\odot$ (left panel) and 18 $M_\odot$ (right panel) MESA models undergoing solid-body rotation at the rate defined by $J_\mathrm{NS}(R_\mathrm{tidal})$ at various stages of post-main sequence evolution. The center panel shows the CE outcome regions detailed in Figure 1, left panel, with dashed lines mapping the increase in radius from bottom to top over the post-main sequence of two stellar models. Outside panels show the normalized specific angular momentum profile of the core through the evolution tracks shown, with line color matching the corresponding region of CE outcome. The minimum specific angular momentum value required for disk formation is indicated by the blue dotted line, and the minimum core mass required to collapse the NS to BH is indicated by the vertical gray line. Note that in every case in which merger occurs (brown and coral), the cores undergo total disk formation with the most conservative assumptions about rotation. Partial disk formation would only be possible when merging with extended models (beige), but CE ejection prevents such mergers from occurring.}
    \label{fig:combo}
\end{figure*}

In Figure \ref{fig:combo}, we show a representative sample in the NS case of how the angular momentum content of the stellar core differs as the primary evolves through the giant branch and the core becomes more compact. The center panel shows a schematic of the evolution of an $8\ M_\odot$ star (A$\rightarrow$B) and an $18\ M_\odot$ star (C$\rightarrow$D) through the mass-radius parameter space shown in the left panel of Figure \ref{fig:CE}, with time proceeding bottom to top. These stellar models are selected as representative of mergers with disrupted and non-disrupted cores, respectively, while the $18\ M_\odot$ model is the same used for analysis in Figures \ref{fig:3D} and \ref{fig:jcomp}, as well as in the simulations of \citet{Tenley}. 

The outer panels of Figure \ref{fig:combo} show the specific angular momentum content of the core material, normalized by $j_\mathrm{isco}$, as a function of mass for the series of models represented in the center panel, with time proceeding top to bottom, calculated with MESA profiles in the same manner we describe above. The evolution depicted in the outer panels is representative of the evolution of all other models in the same regimes (disruptive or non-disruptive mergers). Line color corresponds to the region in the center panel that the stellar model occupies, such that all beige models are excluded from forming \TZOs\ due to envelope ejection prior to merger.

In order to form something like a classical T\.{Z}O, a stable halo structure would need to form around the NS star and persist for some length of time, supporting energy generation via quasi-spherical accretion (in the lower mass case) and/or spherically symmetric nuclear burning at the base of the envelope (in the higher mass case). In Figure \ref{fig:combo}, this would appear as a model for which some central region of the core maintains $j/j_\mathrm{isco}$ values below unity, depicted by the blue dotted line, in which partial-disk formation might occur in the outer region of the core while the interior establishes a slow-rotating core-halo structure. The only cores which would satisfy such partial-disk formation in Figure \ref{fig:combo} are CE ejection cases (beige). In fact, when we apply this analysis across the full parameter space of models ($5-40\, M_\odot$), we find that in every binary system that fulfills the energetic criterion for merger (coral or brown), the core is rotating well above this limit based solely on the orbital angular momentum of the NS at $R_\mathrm{core}$. 

In all so-called \TZO\ progenitor systems, the core must undergo total disk formation upon merger. This is further supported by the FLASH simulations (Figure \ref{fig:jcomp}), in which the shockwaves from inspiral spin up the core such that its angular momentum content satisfies the criterion for total disk formation when the secondary is still as much as $5 R_\mathrm{tidal}$ away from the center of mass and has not yet merged with the core.

This should not be surprising: recent work on the collapse of single giant stars has shown that disk formation is difficult to avoid \citep{2020ApJ...901L..24M}, and that even with zero net angular momentum, convection in the extended hydrogen envelope leads to specific angular momentum profiles greater than $j_\mathrm{isco}$ \citep{2019MNRAS.485L..83Q,  2022MNRAS.511..176A}, leading to accretion disk formation. Though mergers tend to happen when envelopes are more compact, the deposition throughout the primary of orbital angular momentum via shocks guarantees a non-zero net angular momentum even in the innermost core material.

\section{Discussion} \label{sec:discussion}
The above analysis suggests that it is not accretion feedback nor the accompanying jets during inspiral that prevent the formation of T\.{Z}Os, but the immediate formation of an accretion disk upon merger, disallowing the establishment of a stable core-halo structure and leading to either the ejection \citep{Tenley} or collapse \citep{1995MNRAS.274..485P} of the envelope in a matter of years. Consideration of the most basic details of CE inspiral thus precludes the formation of classical \TZOs\ from field binaries altogether.

This is not to say that so-called \TZO\ progenitors are not of great interest. Such systems may be the precursors to a broad range of transients occurring in succession in the same system: starting as an X-ray binary, then producing a recombination transient through partial envelope ejection during CE inspiral \citep[i.e. luminous red nova; see, e.g.,][]{2017ApJ...835..282M,2022ApJ...937...96M}, followed by  collapse to a black hole and a subsequent accretion feedback transient \citep[i.e., gamma-ray burst;][]{1999ApJ...524..262M,2001ApJ...550..410M,2001ApJ...550..357Z,2004MNRAS.348.1215I}. The effects of the unique post-merger mass distribution and morphology of these systems may even lend them to new types of transients, such as the supernova-like merger-driven explosions of \citet{2020ApJ...892...13S} or ultra-long Gamma-ray bursts of \citet{Tenley}, that could be detected by surveys such as Rubin/LSST \citep{2019ApJ...873..111I} and Swift \citep[Section 6]{2009ARA&A..47..567G}.

But perhaps the story of \TZOs\ need not end here. After the dynamic process of merger and the NS's collapse to BH, we speculate that an alternative steady-state merger product may be possible: a thin-envelope \TZO\ (TET\.{Z}O). 

\subsection{Reimagining \TZOs}\label{subsec:tetzo}
\begin{figure}[tbp]
    \figurenum{7}
    \epsscale{1.15}
    \plotone{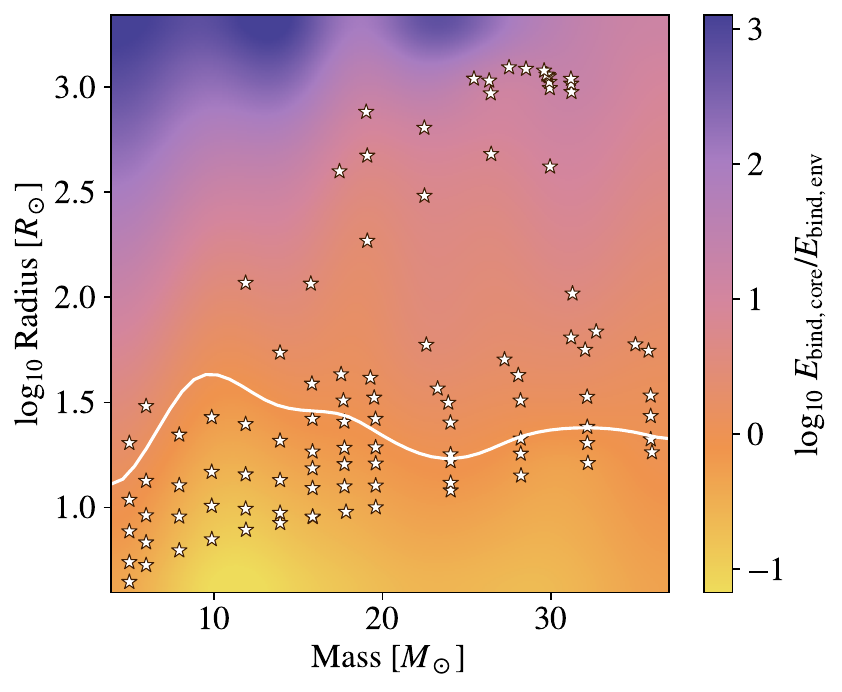}
    \caption{NS merger models overplotted as white stars on mapped values of the ratio of the gravitational binding energy held in the stellar core versus that held in the envelope. In order to retain a thin envelope in a merger scenario, the energy deposition due to the accretion of the core should be insufficient to unbind the remaining envelope (roughly, $E_\mathrm{bind,core}/E_\mathrm{bind,env}<1$, delimited by the white line). Using this ratio as a rough estimate, this criterion is satisfied only for stellar companions in the early stages of the post-main sequence.}
    \label{fig:Ebind}
\end{figure}

Some fraction of the envelope will be ejected during CE inspiral, but due to accretion feedback, any remaining envelope will be quickly unbound in most cases. However, in cases where the core's binding energy is not as dominant when compared to that of the envelope,  a thin-envelope may be retained close to the core if the efficiency of feedback is sufficiently low. The ratio of the binding energy of the core to the binding energy of the envelope is shown for the NS case in Figure \ref{fig:Ebind}. Only in the the late stages of the main sequence and the earliest stages of the post-main sequence is this ratio at or below unity, suggesting that TE\TZO\ candidates are most likely formed from close binaries in which the stellar partner has not evolved far from the main sequence, as we see in the forward-evolved model of LMC X-4 from \citet{Tenley}.

In comparison to wider binaries at the onset of merger, such systems contain less total angular momentum and orbital energy while having a more tightly bound envelope, all of which serve to increase the likelihood some envelope may remain once the merger and subsequent collapse are complete. A steady-state TE\TZO\ occurs when the envelope remnant, supported by radiation pressure, is able to achieve hydrostatic equilibrium about the accreting BH. The timescale from merger to TE\TZO\ formation is unclear due to various uncertainties in mass ejection and the intervening transient events noted above; these will be explored in detail in a follow-up paper.

\begin{figure}[tbp]
    \figurenum{8}
    \epsscale{1.15}
    \plotone{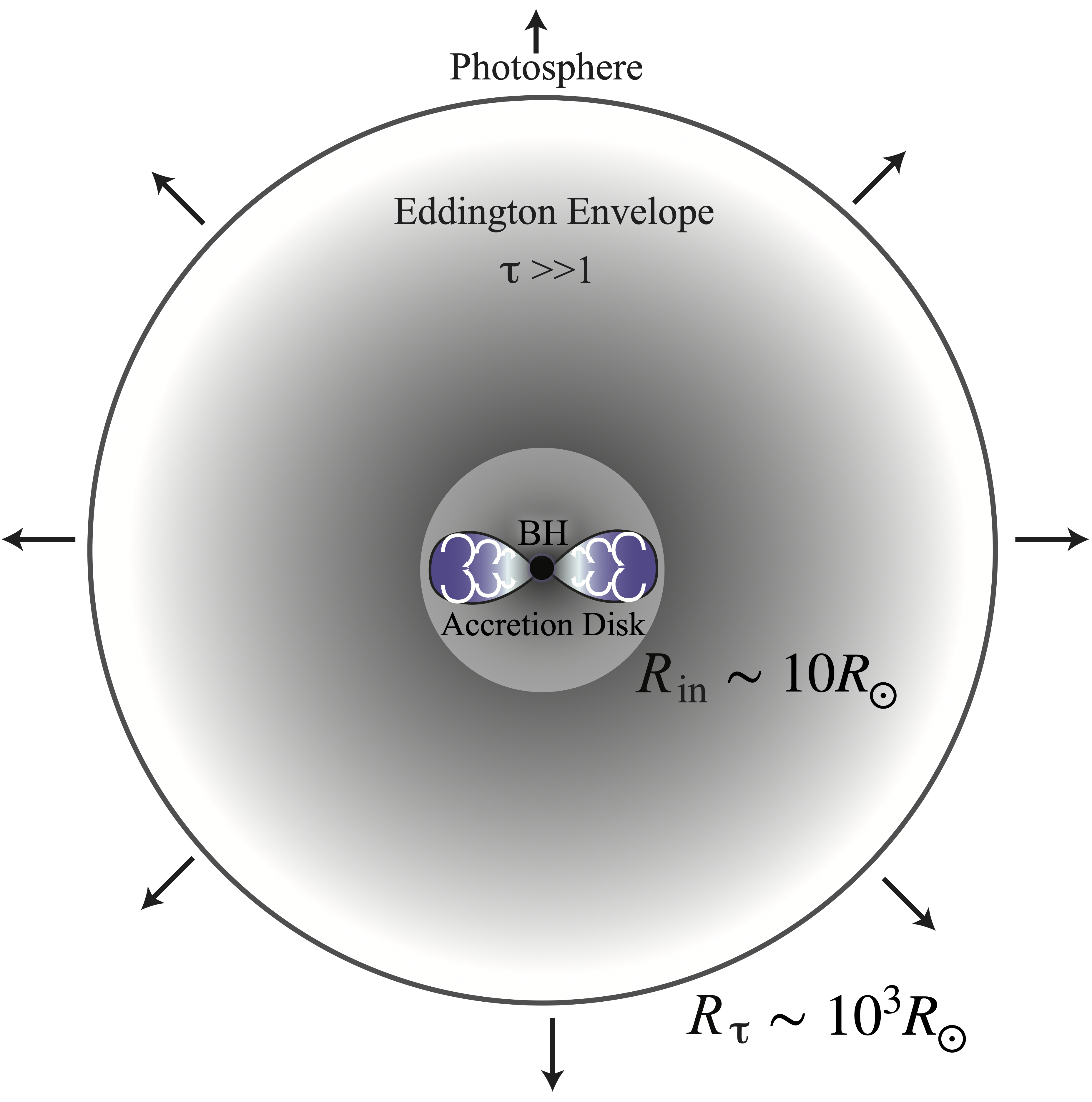}
    \caption{A schematic of the structure of a TE\TZO\ is shown, illustrating a possible configuration for merger remnants following several transient events. After most of the stellar envelope has been ejected or depleted and the NS has collapsed due to mass accretion, a central BH of a few solar masses accretes via disk (with a typical circularization radius $R_\mathrm{in}$) with the disk providing radiation support to an optically-thick thin-envelope that extends to the photosphere at $R_\mathrm{\tau}$. This radius is calculated where the radiation-dominated diffuse envelope transitions from being optically thick to optically thin, with emission from the photosphere represented by black arrows.
    }
    \label{fig:tetzo}
\end{figure}

A schematic of the general structure of a TE\TZO\ is shown in Figure \ref{fig:tetzo}. We consider a radiation supported envelope with negligible self gravity, with less than $1\%$ of the initial envelope mass remaining with the rest being  ejected by accretion feedback that gave rise to the preceding bright transient. The thin-envelope is feeding matter to the  black hole, converting a fraction $\epsilon$  of the accreting mass into radiation ($L=\epsilon \dot{M}c^2$), which is then reprocessed through the optically thick thin-envelope before escaping. The conversion of envelope mass into radiation in the accretion disk supports the thin-envelope in much the same way as originally envisioned by Thorne and \.{Z}ytkow \citep{1975ApJ...199L..19T,1977ApJ...212..832T}, in which gravity and radiation pressure provide a self-regulating mechanism that allows the envelope of the TE\TZO\ to approach a steady state. 

There are several examples in the literature of hypothetical configurations of this type, i.e. envelopes powered by accreting black holes, including the tidal disruption event (TDE) remnants described by \citet{Loeb_1997} and  quasi-stars \citep{2008MNRAS.387.1649B,2011MNRAS.414.2751B,2012MNRAS.421.2713B}. The latter are proposed as possible seeds for supermassive black holes, forming after a massive Population III star ($\sim 10^4 M_\odot$) undergoes core-collapse while retaining a massive ($\gtrsim 10^3 M_\odot$) convective envelope. Though quasi-stars may initially contain a stellar mass black hole, like that in a TE\TZO, their envelopes are three orders of magnitude more massive than their central black holes. Such a configuration is not consistent with a TE\TZO, which will have ejected nearly all envelope material ($M_\mathrm{env} \ll M_\mathrm{BH}$) and have a gravitational potential dominated by the central black hole rather than the remnant envelope. The TDE remnants of \citet{Loeb_1997} are powered by the accretion of debris from a disrupted star onto a massive black hole ($\sim 10^6 M_\odot$), mirroring the central mass-dominated gravitational potential and diffuse envelope of a TE\TZO, as well as the requisite disk accretion in the central engine. The scale and formation channel of this configuration significantly depart from that of a TE\TZO, but the differences in structure that result (e.g., the inner radius of the envelope in a TE\TZO is not defined by tidal disruption, but its exact value impacts the observable properties only logarithmically, therefore minimally) mainly impact the the effective temperature and lifetime, as we will demonstrate.

The stabilizing feedback provided by gravity and radiation pressure in a TE\TZO\ occurs as a result of the following processes. An increase in the luminosity above the Eddington limit, 
\begin{eqnarray} \label{eqn:11}
L=L_{\rm Edd}  &=& { 4\pi G \mu_e m_p c M_{\rm BH} \over \sigma_\mathrm{T}} \nonumber \\
 &=& 1.4\times 10^{39} \left( {M_{\rm BH} \over 10\, M_\odot} \right)\;{\rm erg/s},
\end{eqnarray}
would result in an outflow and, as a result, reduce the accretion luminosity. Here $\mu_e$  is the mean atomic weight per electron, $m_p$ is the proton mass and $\sigma_\mathrm{T}$ is the Thomson scattering cross-section. Conservation of momentum demands that mass is ejected from the photosphere of the envelope as a wind, $\dot{M}_{\text{wind}} v_\infty \approx L_{\text{Edd}} / c$, where $v_\infty$ is the escape velocity at the photosphere ($R_{\tau}$) as defined by $\tau=1$. Accretion below the Eddington limit would, on the other hand, allow the infall rate to increase, which would return the luminosity to its equilibrium value \citep{1978ApJ...226.1041C}. 

The prior envelope ejection will carry away most of the angular momentum from the outer regions, leaving any remnant thin-envelope very slowly rotating. Thus we do not expect rotation to impact the morphology of what we assume to be a spherical, steady state, optically thick thin-envelope surrounding the black hole as our expected appearance for a TE\TZO.

The effective temperature at the photosphere can be expressed in terms of the luminosity and the extent, hence the mass, of the thin-envelope \citep{Loeb_1997}: 
\begin{eqnarray} \label{eqn:12}
T_{\rm ph} &=& \left({L_{\rm Edd} \over 4 \pi  R_{\tau}^2 \sigma}\right)^{1/4} \nonumber \\
&\approx&  10^{6} \left( {M_{\rm BH} \over 10\, M_\odot} \right)^{1/4}  \left( \frac{M_{\text{env}}}{5 \times10^{-3}\, M_\odot}   \right)^{-1/4}  {\rm K} ,
\end{eqnarray}
in which $\sigma$ is the Stefan-Boltzmann constant.
The effective temperature has a weak dependence on $M_{\rm BH}$ and $M_{\text{env}}$, though with more sensitivity to small changes in the thin-envelope mass, and corresponds to a blackbody spectrum peaking in the soft X-ray band or, should $\gtrsim 1 \%$ of the initial envelope mass remain, peaking instead at UV energies. For a constant radiative efficiency ($\epsilon= 0.1\epsilon_{-1}$), the lifetime of a TE\TZO\ can  thus be written as  
\begin{equation} \label{eqn:13}
t_{\rm life}\approx 10^4 \left( {M_{\rm BH} \over 10\, M_\odot} \right)^{-1}  \left( \frac{M_{\text{env}}}{5 \times10^{-3}\, M_\odot}   \right) \epsilon_{-1}  \;{\rm yr}. 
\label{life}
\end{equation}
The above envelope configuration  could exist for as long as $\approx 10^4-10^6$ years, depending on $M_{\rm BH}$, $\epsilon$, and $M_{\text{env}}$. 

We can then estimate how many TE\TZOs\ are predicted to reside in galaxies like our own. Using the CE population synthesis results from \citet{2023A&A...672A..54S}, based on the \texttt{Fiducial} model parameters in COMPAS from \citet{2018MNRAS.481.4009V} and detailed CE outcomes developed by \citet{2020PASA...37...38V}, we expect a merger rate of TE\TZO-type progenitor binaries comprised of a NS and massive companion in Milky Way-type galaxies \citep[with star formation rate $\sim 8 M_\odot$ yr$^{-1}$; see][]{2023A&A...672A..54S} to be $\sim 10^{-3}$ yr$^{-1}$. With a lifetime of $\approx 10^4$ years (Equation \ref{life}),  we thus expect a handful of candidates per galaxy if post-merger TE\TZO\ formation is typical of these systems.

\subsection{In Search of TE\TZOs}\label{subsec:ULX}
Interestingly, TE\TZOs\ have predicted luminosities ($\approx10^{39}$ erg/s) and photon temperatures ($\approx0.1 $ keV) that are similar to ultraluminous X-ray sources (ULX) and should occur at similar rates of a few per galaxy \citep{2024A&A...681A..16T}. ULXs preferentially appear in regions that have recently undergone high rates of star formation, as do X-ray binaries and massive stars \citep{2006ApJ...642..171L,2012AJ....144...12S,2019MNRAS.483.5554E}, 
therefore we speculate that ideal TE\TZO\ progenitors share the same environment as ULX progenitors. Though it is still an open question whether NSs or BHs are the dominant engines of ULXs \citep{2021A&A...649A.104G,2022MNRAS.509.1587W}, there is high-quality data supporting that some ULX properties may be best explained by accreting stellar-mass BHs \citep{2018MNRAS.479.4271P}. Therefore it is plausible that TE\TZOs\ may not only be the end products of CE events that lead to the merger of X-ray binaries, but that they may have already been uncovered.

One indication that a ULX may be associated with a TE\TZO\ could come from high-cadence time-domain surveys such as LSST \citep{2019ApJ...873..111I} and eRosita \citep{2020NatAs...4..634M}: the formation of a TE\TZO\ would begin with the detection of a bright transient due to the central BH accreting the dense remnant core from the primary. A fairly bright and fairly isotropic  optical transient is expected to accompany the  disruption of the envelope  \citep{2020ApJ...892...13S} while an ultra-long gamma-ray burst \citep{Tenley} might be detected for observers along the axis of the jet. Rather than fading away, the remnant would instead settle down over time to a steady-state ULX. An event of this kind would need to be relatively close for the post-merger ULX to be seen, but could be a clear signature indicating TE\TZO\ formation. 

\section{Summary} \label{sec:conclusions}

In this paper, we set out to find the formation pathways that would lead to \TZO\ formation from field binaries. After constraining our parameter space of progenitors through the lens of common envelope ejection criteria, we have explored the implications of angular momentum deposition on the primary's core based on global properties of the star and its companion (e.g. $M_\mathrm{core}$, $R_\mathrm{core}$, $M_\mathrm{CO}$), 3D hydrodynamics based on the setup of \citet{Tenley}, and analysis of a library of detailed 1D stellar models. Upon ruling out the formation of \TZOs\ from these systems, we have proposed other possible outcomes based on our findings. So-called \TZO\ progenitors are of great interest to transient astronomy due to their potential to sequentially produce a broad range of transient phenomena across the electromagnetic spectrum, and further work to understand the varied and dynamic lifetimes of these systems is merited.

The key conclusions of this work are the following:

\begin{itemize}
    \item Classical T\.{Z}Os are unlikely to be assembled via common envelope interactions in field binaries. The merger process favors the conditions required to form an accretion disk in the core, which prevent the radial accretion that would power lower mass \TZOs\ and initiate envelope collapse in higher mass \TZOs. The core structure of the primary during merger has no bearing on \TZO\ formation, regardless of whether it remains intact or is disrupted.
    \item We propose an alternative merger product that may form around the resulting stellar-mass black hole after collapse: the thin-envelope T\.{Z}O (TET\.{Z}O). Feedback from the accretion of the remaining core material is likely to eject most or all of the stellar envelope, but if a minimal amount ($\lesssim1\%$) remains, a steady-state configuration may arise in which the optically thick thin-envelope is powered by the accretion luminosity of the black hole.
    \item We find that post-merger TE\TZOs\ may plausibly be associated with ultra-luminous soft X-ray sources and that they may have already been detected in nearby galaxies at rates that are comparable with those predicted here.
\end{itemize}

\begin{acknowledgments}
We gratefully acknowledge A. Batta, D. Coulter, A. Grichener, A. Kolborg, M. Renzo, S. Schr{\o}der, and R. Yarza for helpful discussions. We also sincerely thank the anonymous referee for their insightful feedback, which has significantly clarified and strengthened this work. R.W.E. acknowledges the support of the University of California President's Dissertation-Year and Eugene V. Cota-Robles Fellowships, the Heising-Simons Foundation, the ARCS Foundation, and the Vera Rubin Presidential Chair for Diversity at UCSC. This material is based upon work supported by the National Science Foundation Graduate Research Fellowship Program under Grant No. \texttt{1339067}. T.H.-S. acknowledges the support of the University of California - Historically Black College and University (UC-HBCU) Fellowship. A.V.-G. acknowledges funding from the Netherlands Organisation for Scientific Research (NWO), as part of the Vidi research program BinWaves (project number 639.042.728, PI: de Mink). E.R.-R. acknowledges support from the Heising-Simons Foundation and the National Science Foundation (Grant Nos. \texttt{2150255} and \texttt{2307710}). Any opinions, findings, and conclusions or recommendations expressed in this material are those of the authors and do not necessarily reflect the views of the NSF. The 3D hydrodynamics software used in this work was developed in part by the DOE NNSA- and DOE Office of Science-supported Flash Center for Computational Science at the University of Chicago and the University of Rochester. Simulations in this paper made use of the COMPAS rapid binary population synthesis code (version 02.13.01), which is freely available at \href{http://github.com/TeamCOMPAS/COMPAS}{http://github.com/TeamCOMPAS/COMPAS}.
\end{acknowledgments}

\vspace{5mm}
\software{\texttt{Python},
	\texttt{MESA} \citep{Paxton2011, Paxton2013, Paxton2015, Paxton2018, Paxton2019}, 
 \texttt{FLASH} \citep{2000ApJS..131..273F},
 \texttt{COMPAS} \citep{2022ApJS..258...34R},
	\texttt{matplotlib} \citep{Hunter2007}, 
 \texttt{yt} \citep{2011ApJS..192....9T},
	\texttt{NumPy} \citep{vanderwalt2011}, 
	\texttt{py\_mesa\_reader} \citep{WolfSchwab2017}}

\bibliography{refs}{}
\bibliographystyle{aasjournal}

\end{document}